\documentstyle[graphicx,aps]{revtex}

\newcommand{\ybco} {YBa$_2$Cu$_3$O$_7$ }
\newcommand{\sto} {SrTiO$_3$ }

\begin{document}


\title{Ratchet effect in dc SQUIDs}

\author{\renewcommand{\thefootnote}{\alph{footnote})} S.~Weiss,
D.~Koelle\footnote{\vspace*{-0cm}e-mail: koelle@ph2.uni-koeln.de}, J.~M\"{u}ller, and R.~Gross}
\address{II.~Physikalisches Institut, Universit\"{a}t zu K\"{o}ln, D-50937 K\"{o}ln, Germany }
\author{K. Barthel}
\address{Institut f\"{u}r Schicht- und Ionentechnik, Forschungszentrum J\"{u}lich, D-52425 J\"{u}lich, Germany }

\maketitle


\begin{abstract}
We analyzed voltage rectification for dc SQUIDs biased with ac
current with zero mean value. We demonstrate that the reflection
symmetry in the 2-dimensional SQUID potential is broken by an
applied flux and with appropriate asymmetries in the dc SQUID.
Depending on the type of asymmetry, we obtain a rocking or a
simultaneously rocking and flashing ratchet, the latter showing
multiple sign reversals in the mean voltage with increasing
amplitude of the ac current. Our experimental results are in
agreement with numerical solutions of the Langevin equations for
the asymmetric dc SQUID.

\end{abstract}

\vspace*{1cm}




Directed molecular motion in the absence of a directed net driving
force or a temperature gradient in biological systems has drawn
much attention to the so-called Brownian motors
\cite{Haenggi96,Astumian97} where nonequilibrium fluctuations can
induce a net drift of particles along periodic structures which
lack reflection symmetry. Such structures are denoted as ratchets.

In general, a ratchet system is characterized by the equation of
motion of an overdamped particle in a periodic potential $w(x,t)$
\begin{equation}\label{equmotion}\xi\frac{dx}{dt}=-\frac{\partial}
{\partial x}[w(x,t)-F_d(t)x]+F_n(t)\equiv-\frac{\partial}{\partial
x} u(x,t)+F_n(t)
\end{equation}
with $w(x,t) \neq w(-x,t)$ for all choices of origin. Here, $\xi$
is a friction coefficient, $x$ a cyclic coordinate, $F_d(t)$ a
driving force and $F_n(t)$ represents a force with zero mean value
due to thermal noise with Gaussian distribution. One can classify
different types of ratchet systems depending on the actual form of
$w(x,t)$ and $F_d(t)$ (see e.g. \cite{Juelicher97}). The {\it
rocked thermal ratchet} \cite{Magnasco93,Bartussek94} is obtained
if one chooses $w(x,t)$ to be time independent and $F_d(t)$ to be
either stochastic or deterministic. For the {\it flashing ratchet}
a time dependent potential $w(x,t)$ is chosen which, in the
simplest case, is of the form $w(x,t)=w_0(x)w_d(t)$, where
$w_d(t)$ again can be stochastic or deterministic. The common
feature of all ratchet systems is their rectifying  property: it
is possible to extract directed motion from nonequilibrium
fluctuations or periodic excitations with vanishing time average.
Although considerable theoretical work exists in particular for
stochastically driven ratchets, only few experiments have been
realized e.g. by means of dielectric \cite{Rousselet94} and
optical  potentials \cite{Faucheux95}.

The periodicity of the Josephson coupling energy $E_J$ with
respect to the phase difference $\delta$ of the macroscopic wave
function across a Josephson junction is an ideal prerequisite to
build a ratchet system. Within the resistively and capacitively
shunted junction model the equation of motion of $\delta$ is
equivalent to the motion of a particle in the so-called tilted
washboard potential
$U(\delta,t)=-\frac{\Phi_0}{2\pi}\{I_0\cos\delta+I(t)\delta\}$
(see e.g. \cite{Likharev86}). Here, $\Phi_0$ is the flux quantum
and $I_0$ is the maximum Josephson current across the junction. An
external force causing the tilt is obtained by a bias current $I$,
and the mass and friction coefficient are determined by the
junction capacitance $C$ and the junction resistance $R$,
respectively. In the strongly overdamped limit ($\beta_C\equiv
\frac{2\pi}{\Phi_0}I_0R^2C\ll 1$) $C$ can be neglected. The
voltage $V=\frac{d \delta}{d t}\frac{\Phi_0}{2\pi}$ across the
junction represents the velocity of the particle via the second
Josephson equation. For a short Josephson junction with a current
phase relation which is an odd function, $U(\delta)$ is always an
even function and no ratchet potential is obtained. However,
voltage rectification may be induced by asymmetric fluctuations as
proposed in \cite{Zapata98}. To obtain a ratchet potential for
Josephson junction systems, two ratchet configurations based on
{\it coupled} Josephson junctions have been proposed. Zapata et
al. \cite{Zapata96} considered an asymmetric three junction dc
SQUID, with vanishing loop inductance $L$ thereby coupling the
phase differences across the junctions rigidly via the applied
flux $\Phi_a$. Due to the doubled phase shift in the SQUID arm
with the two junctions in series the reflection symmetry in the
effective 1-dimensional potential is broken. Falo et al.
\cite{Falo99} proposed a 1-dimensional Josephson junction array
with spatially alternating critical junction currents and
alternating loop inductances. They showed that a Josephson fluxon
experiences a ratchet potential as it moves along the array. This
effect has been confirmed experimentally very recently
\cite{Trias00}.

In this letter we show that the well known voltage rectification
in asymmetric dc SQUIDs \cite{Waele67} is a consequence of broken
symmetry in the two dimensional SQUID potential due to this
asymmetries and the application of magnetic flux. In contrast to
the model proposed by Zapata et al. \cite{Zapata96} this results
in a quasi two dimensional ratchet potential. We demonstrate its
experimental realization by using a two junction high transition
temperature ($T_c$) dc SQUID with small but finite $L$. In our
experiments we use an asymmetry $\alpha_I$ in the critical
currents $I_0^1=(1-\alpha_I)I_0$ and $I_0^2=(1+\alpha_I)I_0$ of
the two junctions and/or an asymmetry $\alpha_L$ which introduces
a self-field effect, i.e. a flux component
$\Phi_{self}\equiv\alpha_L \Phi_0 I/I_c$ with $I_c\equiv
I_0^1+I_0^2$. A finite value of $\alpha_L$ can be realized for
example with different inductances in the left and the right arm
of the SQUID loop; our definition however is more general and
accounts for all kinds of self field generation.



The extension of the tilted washboard potential model in two
dimensions has been used to provide insight into the dynamics of
both, symmetric dc SQUIDs \cite{Ryhaenen89} and dc SQUIDs with an
$I_0$ asymmetry \cite{Early94}. The SQUID potential $U_s$
\begin{equation}
\begin{array}{ll}
U_s(\delta_1,\delta_2,t)=-\frac{\Phi_0}{2\pi}\{ I_0^1\cos\delta_1
+I_0^2\cos\delta_2\\-\frac{I_c}{4\pi\beta_L}[\delta_2-\delta_1-
2\pi(\frac{\Phi_a}{\Phi_0}+\alpha_L \frac{I(t)}{I_c})
]^2+\frac{I(t)}{2}(\delta_1+\delta_2)\}
\end{array}
\end{equation}
contains three basic terms, (i) the two cosine terms representing
$E_J$, (ii) the quadratic magnetic energy term and (iii) the
external force term proportional to $I$. Here, $\delta_1,
\delta_2$ are the phase differences across the two Josephson
junctions and $\beta_L\equiv I_cL/\Phi_0$ is the normalized
inductance. An asymmetry $\alpha_I \neq 0$ introduces a difference
in $E_J$ for the two junctions thus deforming the double cosine
part of the potential landscape, while $\alpha_L\neq 0$ shifts the
minimum of the parabolic magnetic energy term proportional to $I$.
Transforming into more appropriate phase coordinates along the
direction of the bias current $\delta\equiv(\delta_1+\delta_2)/2$
and the applied flux $\varphi\equiv(\delta_2-\delta_1)/2$ and
using  $i\equiv I/I_c$ and $f_a\equiv \frac{\Phi_a}{\Phi_0}$
yields in analogy to eq.~(\ref{equmotion})
$u_s(\delta,\varphi,t)=w_s(\delta,\varphi,t)-i(t)\delta $ with
\begin{equation}\label{trans.pot2}
\begin{array}{ll}
w_s(\delta,\varphi,t)=&-\cos\delta\cos\varphi+\alpha_I\sin\delta\sin\varphi
\\&+\frac{1}{\pi\beta_L}\{\varphi -\pi f_a-\pi\alpha_Li(t)\}^2.
\end{array}
\end{equation}
In this coordinates, the voltage $V$ across the SQUID is
determined by $\frac{d\delta}{d t}$. A ratchet effect, i.e.
voltage rectification for an external excitation with zero mean
value, will occur if the system is not invariant under parity
transformations with respect to $\delta$, that is for
\begin{equation}\begin{array}{rlr}
w_s(\delta_0+\delta,\varphi_0+\varphi)&\neq
w_s(\delta_0-\delta,\varphi_0+\varphi)& \mbox{and}\\
w_s(\delta_0+\delta,\varphi_0+\varphi)&\neq
w_s(\delta_0-\delta,\varphi_0-\varphi) ,&
\end{array}\end{equation}
for arbitrary origin $(\delta_0,\varphi_0)$. Eq.(\ref{trans.pot2})
satisfies these conditions, i.e. $w_s$ is a ratchet potential in
$\delta$, if $f_a\neq \frac{n}{2}\wedge(\alpha_I\neq
0\vee\alpha_L\neq 0)
;(n= 0,1,2...)$. For given $\alpha_I, \alpha_L$, the applied flux $f_a$ is an
experimentally controllable parameter, which allows to change the
parity of $w_s$. The type of asymmetry of the dc SQUID driven by a
fluctuating or periodic bias current $i(t)$ determines the nature
of our ratchet system. For $\alpha_I\neq 0, \alpha_L=0$, $w_s$ is
time independent, thus we have a rocking ratchet system. For
$\alpha_L\neq0$, $w_s$ becomes time dependent since the minimum of
the magnetic energy term shifts proportional to $i(t)$. In that
case we obtain a simultaneously flashing and rocking ratchet.

Figure~\ref{potential} shows contour plots of the 2-dimensional
SQUID potential $u_s$ for $\alpha_L=0$ and different values of
$\alpha_I$ and bias current $i$. For $\alpha_I=0$ (upper row), the
potential has reflection symmetry with respect to $\delta$, and
with finite $i$ we have
$u_s(\delta,\varphi;i)=u_s(-\delta,\varphi;-i)$ and thus
$V(i)=-V(-i)$. For $\alpha_I=0.3$ (lower row) no reflection
symmetry is present and in fact, while for $i=+0.9$ no local
minima are found i.e. the solution for the phase is unbounded, for
$i=-0.9$ local minima are present and no voltage appears.



The signature of a ratchet effect can be observed in the
$V(\Phi_a)$ characteristics of an asymmetric SQUID ($\alpha_I\neq
0$) in form of a flux shift $\Delta \Phi$ (see Fig.~\ref{vphi})
between the two curves obtained by biasing the SQUID with the
currents $\pm I$ with an appropriate choice of $|I|\approx I_c$
(experimental details are given below). A flux shift $\Delta \Phi
\neq n\Phi_0$ corresponds to $V(I,\Phi_a) \neq -V(-I,\Phi_a)$ for
all $\Phi_a \neq \frac{n}{2}\Phi_0$ ($n=$0,1,2..). Thus the bias
current (or time-) reversal symmetry is broken, which is true for
every tilted ratchet potential: for certain values of the external
force the magnitude of the velocity of the system depends on the
sign of the force. Note however, that $V(I,\Phi_a) =
-V(-I,-\Phi_a)$ always holds. Applying an ac excitation at angular
frequency $\omega$ with zero mean to a SQUID with $\Delta\Phi\neq
n\Phi_0$ results in a finite mean voltage, which is maximum for
$\Phi_a=\frac{2n+1}{4}\Phi_0$ and zero for
$\Phi_a=\frac{n}{2}\Phi_0$. Obviously the maximum mean voltage
with ac bias is largest if $\Delta\Phi=0.5\Phi_0$.


Next we analyze the dependence of $\Delta \Phi$ on $\alpha_L$,
$\alpha_I$, $\beta_L$ and $I$. With self field effect and
symmetric junctions ($\alpha_L\neq 0$, $\alpha_I=0$), $\Delta
\Phi$ between a pair of $V(\Phi_a;\pm I)$ curves is simply
proportional to the bias current and $\alpha_L$:
$\Delta\Phi=2\Phi_0\alpha_L I/I_c$. For $\alpha_L=0$,
$\alpha_I\neq 0$, $\Delta \Phi$ depends on $\alpha_I$, $\beta_L$
and can be obtained as follows. The flux value $\Phi_1$ (cf.
Fig.~\ref{vphi}) for $I\sim I_c$ is equal to the flux value of
maximum critical current $I_c=I_0^1+I_0^2$. $I_c$ is reached if
the circulating current in the loop equals $\alpha_I I_0$,
corresponding to an applied flux
$\frac{1}{2}\Phi_0\alpha_I\beta_L$. From our definition it follows
that $\Delta\Phi=\Phi_0\alpha_I\beta_L$. For a SQUID with given
inductance this relation can be used to determine the junction
asymmetry $\alpha_I$ by measuring $\Delta\Phi$ and $I_c$. We note
that $\Delta \Phi\neq0$ independent on $I$ is frequently observed
for our high-$T_c$ SQUIDs which is not surprising, since the
spread of the critical currents of the two junctions can be
substantial \cite{Koelle99}.


To obtain more insights into the dynamics of the system we
performed numerical simulations to solve the coupled Langevin
equations of the asymmetric dc SQUID \cite{Kleiner95} with
$\beta_C\ll 1$ and with thermal fluctuations due to Nyquist noise
in the junction resistors which adds white noise with zero mean
and Gaussian distribution. We use the normalized noise parameter
$\Gamma\equiv\frac{2\pi k_BT}{\Phi_0I_c/2}$ with thermal energy
$k_BT$. In our simulations we chose the normal resistance of the
two junctions $R_n^1=2R_n/(1-\alpha_I)$, $R_n^2=2R_n/(1+\alpha_I)$
thus keeping the characteristic voltage $V_c\equiv
I_cR_n=I_0^1R_n^1=I_0^2R_n^2$ and the corresponding intrawell
relaxation frequency $\omega_0=\frac{2\pi}{\Phi_0}V_c$ of each
junction constant ($R_n$ is the SQUID normal resistance). This
choice corresponds to an $I_0$ asymmetry due to different junction
areas. We calculated the normalized mean voltage $<v>\equiv
\frac{1}{I_cR_n}\frac{1}{T}\int_0^T V(i(t))dt$ for an ac
excitation $i(t)=i_{ac}\sin(\omega t)$ with normalized amplitude
$i_ {ac}\equiv I_{ac}/I_c$. The integration period $T$ was chosen
to be $T>2\pi/\omega$ (for $\omega<\omega_0$). We first discuss
the {\it rocking ratchet} case ($\alpha_I\neq 0, \alpha_L=0$).
Here, we find $<v>(i_{ac})$ characteristics as shown in
Fig.~\ref{alpha_I}, which are strikingly similar to those
calculated for a 1-dimensional rocked thermal ratchet
\cite{Bartussek94}. For $\Gamma=0$ [Fig.~\ref{alpha_I}~(a)] we
find for adiabatically slow excitation, $\hat{\omega}=0.001$
($\hat{\omega}\equiv \omega/\omega_0$), a lower ($i_l$) and an
upper ($i_u$) threshold for $i_{ac}$. For $i_{ac}<i_l$, no mean
voltage appears. For $i_l<i_{ac}<i_u$ the mean voltage increases
monotonically with increasing $i_{ac}$ and for $i_{ac}>i_u$ the
voltage decreases monotonically. For $\hat{\omega}=0.01$, the
overall shape of $<v>(i_{ac})$ is similar to the
$\hat{\omega}\rightarrow 0$ case, however, voltage steps at
$<v>=n\hat{\omega}$ appear. These steps can be interpreted as
Shapiro steps \cite{Likharev86}, where the phase dynamics
synchronizes with the external excitation. For a given $i_{ac}$
and within one excitation period the phase "rolls" $m$2$\pi$ in
one current direction and $k$2$\pi$ in the opposite current
direction ($m,k$: integer). On average the phase evolves
$(m-k)$2$\pi$ per excitation period and hence
$<v>=(m-k)\hat{\omega}$. Within $i_l<i_{ac}<i_u$, we have $k=0$
and $m$ increases with increasing $i_{ac}$. For $i_ {ac}>i_u$
both, $m$ and $k$ increase with increasing $i_{ac}$, while $(m-k)$
exhibits a switching behavior [see inset Fig.~\ref{alpha_I}~(a)]
with an overall decrease with increasing $i_{ac}$. As
$i_{ac}\rightarrow\infty$ we have $(m-k)\rightarrow0$. With
increasing frequency the value of $(m-k)$ decreases. For
$\hat{\omega}=0.1$ a finite voltage appears in certain
$i_{ac}$-intervals only, where plateaus of equal voltage appear,
i.e. $(m-k)=0,1$. A further increase of $\hat{\omega}$ confines
the locking condition to obtain $(m-k)>0$ to smaller
$i_{ac}$-intervals. For $\hat{\omega}=0.5$, without noise, no mean
voltage is predicted. If one introduces small thermal fluctuations
($\Gamma=0.03)$ [see Fig.~\ref{alpha_I}(b)], the condition for
$(m-k)>0$ is relaxed and a stochastic resonance like effect sets
in (for a recent review see \cite{Gammaitoni98}). While the
maximum voltage for each frequency is reduced the sharp plateaus
observed for $\Gamma=0$ smear out. Furthermore, even for
$\hat{\omega}=0.5$ there exists an $i_{ac}$-range wherein
$<v>\neq0$.

We next consider the {\it flashing ratchet} case ($\alpha_L\neq
0$). Here, a novel feature appears: the mean voltage undergoes
multiple sign reversals with increasing amplitude $i_{ac}$ as
shown in Fig.~\ref{alpha_L}. The envelope of the peak voltages
exhibits an oscillatory damped behavior, and with finite
$\Gamma=0.03$ the step like behavior is again smeared out. An
additional $I_0$-asymmetry [Fig.~\ref{alpha_L}(b)] further
complicates the $<v>(i_{ac})$-characteristics. For $\Gamma=0$, the
oscillation period of the envelope is clearly smaller as in
Fig.~\ref{alpha_L}(a), however, this oscillation is less obvious
if one adds small thermal fluctuations ($\Gamma=0.03$). In any
case a finite $\Gamma$ induces further sign changes in
$<v>(i_{ac})$.


To test our model we performed measurements on a \ybco thin film
dc SQUID with $24^\circ$ bicrystal Josephson junctions on a \sto
substrate. At $T=$78K, we found $I_c=$172 $\mu$A
($\Gamma\approx$0.04), $R_n=$1.34$\Omega$, $V_c=$232$\mu$V,
$\omega_0/2\pi=$112GHz and $\beta_c\approx 0.07$. From the
estimated geometric inductance we obtain $L\approx 30$pH. However,
we note that due to the unknown contribution from the kinetic
inductance, $L$ may be substantially larger. All measurements were
carried out in a magnetically shielded cryostat. To apply the ac
bias current we used two different techniques: in the adiabatic
frequency regime up to 500kHz the ac current was fed via a twisted
pair of wires to the SQUID. At higher frequency (12.31GHz) the
microwave power was guided by a semi rigid coaxial cable into the
cryostat. At the end of the coaxial cable an antenna structure was
positioned about 1cm above the SQUID. In this geometry the
microwave coupling is not well defined, which implies that both,
in- and out- of phase currents are induced across the two
junctions. The latter corresponds to a circulating current in the
SQUID loop. Hence, a substantial self-field effect can be
generated which is not present in the adiabatic experimental
setup. For details of the experimental setup and SQUID layout see
\cite{Weiss00}.

The measured mean voltage vs. $I_{ac}$ at adiabatically slow
excitation ($\omega/2\pi=50$kHz) is shown in Fig.~\ref{experiment}(a)
together with the numerical simulation result for $L=$60 pH
corresponding to $\beta_L=5.1$. The measured flux shift $\Delta
\Phi\approx 0.5\Phi_0$ (see Fig.~\ref{vphi}) yields a junction
asymmetry of $\alpha_I=\Delta/\Phi_0\beta_L\approx$0.1. The good
agreement is a further confirmation of a ratchet effect in our
system. In contrast to the adiabatic case a much more complex
behaviour occurs at $\omega/2\pi=12.31$GHz ($\hat{\omega}\approx
0.1$) as shown in Fig.~\ref{experiment}(b). The mean voltage
oscillates with increasing ac amplitude and multiple sign
reversals occur. At this frequency the ac bias is not well defined
and the self-field effect can play a major role. In fact, the
qualitative agreement with the simulation results shown in
Fig.~\ref{alpha_L}(b) for $\Gamma=0.03$  suggests that the
assumption of a finite value of $\alpha_L$ is reasonable. We
believe that the observed behaviour can be ascribed to a flashing
rocked ratchet system.

In conclusion, we propose asymmetric dc SQUIDs as a model system
for the study of the ratchet effect. We show that the voltage
rectification effect in dc SQUIDs occurs due to broken reflection
symmetry in the 2-dimensional SQUID potential and present a model
which accounts for this observation by introducing a critical
current asymmetry and a self field effect. The latter, although
found to be responsible for a rich and complex dynamic behaviour,
requires better control of high frequency coupling to the SQUID in
further experiments. For the asymmetry due to different critical
junction currents we obtain a simple rule to design SQUIDs with a
large ratchet effect. It should be pointed out that the observed
ratchet effect is no genuine property of high-$T_c$ SQUIDs. In
fact, using conventional low-$T_c$ SQUIDs might have advantages
due to the superior low spread junction technology. However the
voltage scale on which the ratchet effect occurs is determined by
the $I_cR_n$ product which is larger for high-$T_c$ SQUIDs
\cite{Koelle99}. Finally we emphasize that the observation of a
ratchet effect in asymmetric dc SQUIDs opens up the perspective
for a variety of experimental studies in a system with
straightforward detection of directed motion, simply by measuring
voltage. The system offers experimental control over important
parameters such as amplitude and frequency of the external force,
damping coefficient and thermal noise parameter. Furthermore, the
impact of different types of fluctuations might be interesting to
investigate.

We gratefully acknowledge R.~Kleiner for providing us with his
powerful simulation tool, and we thank T. Bauch and A.
Schadschneider for fruitful discussions.

\pagebreak



\begin{figure}[p]
\vspace*{0.5cm} \noindent Fig.\,1 Contour plot of the
\underline{}SQUID potential $u_s$ ($\beta_L=1.5$, $f_a=0.25$) with
and without $I_0$ asymmetry (upper row: $\alpha_I=0$, lower row:
$\alpha_I=0.3$) and for both directions of bias current $i$.
Arrows indicate direction of motion along phase trajectories
(dashed lines).

\vspace*{.5cm} \noindent Fig.\,2 Measured dc voltage $V$ vs.
applied flux $\Phi_a$ for different values of bias current $I$
with a flux shift $\Delta \Phi=|\Phi_1-\Phi_2|\approx 0.5\Phi_0$,
with $\Phi_{1,2}$ defined as the flux values of corresponding
extrema.

\vspace*{.5cm} \noindent Fig.\,3 Calculated mean voltage $<v>$
across dc SQUID with $I_0$ asymmetry ($\alpha_I=0.3$) vs. ac
amplitude $i_{ac}$ without thermal noise (a) and for $\Gamma=0.03$
(b) for different frequencies and $f_a=0.25$. The inset in (a)
shows an expanded view for $\hat{\omega}=0.01$ at
$i_{ac}\approx$1. The labels for $\hat{\omega}$ in (b) are also
valid for (a).

\vspace*{.5cm} \noindent Fig.\,4 Calculated mean voltage $<v>$
across dc SQUID with self-field effect vs. ac amplitude $i_{ac}$
without thermal fluctuations (thick line) and for $\Gamma=0.03$
(dots); $f_a=0.25$.

\vspace*{.5cm} \noindent Fig.\,5 Mean voltage across dc SQUID vs.
ac amplitude at $f_a=0.25$:
(a) comparison of experiment and calculation in the adiabatic limit;
(b) experiment for $\hat{\omega}\approx0.1$.

\end{figure}


\pagebreak \thispagestyle{empty}
\begin{figure}[p]
\center{\includegraphics [width=0.5\columnwidth] {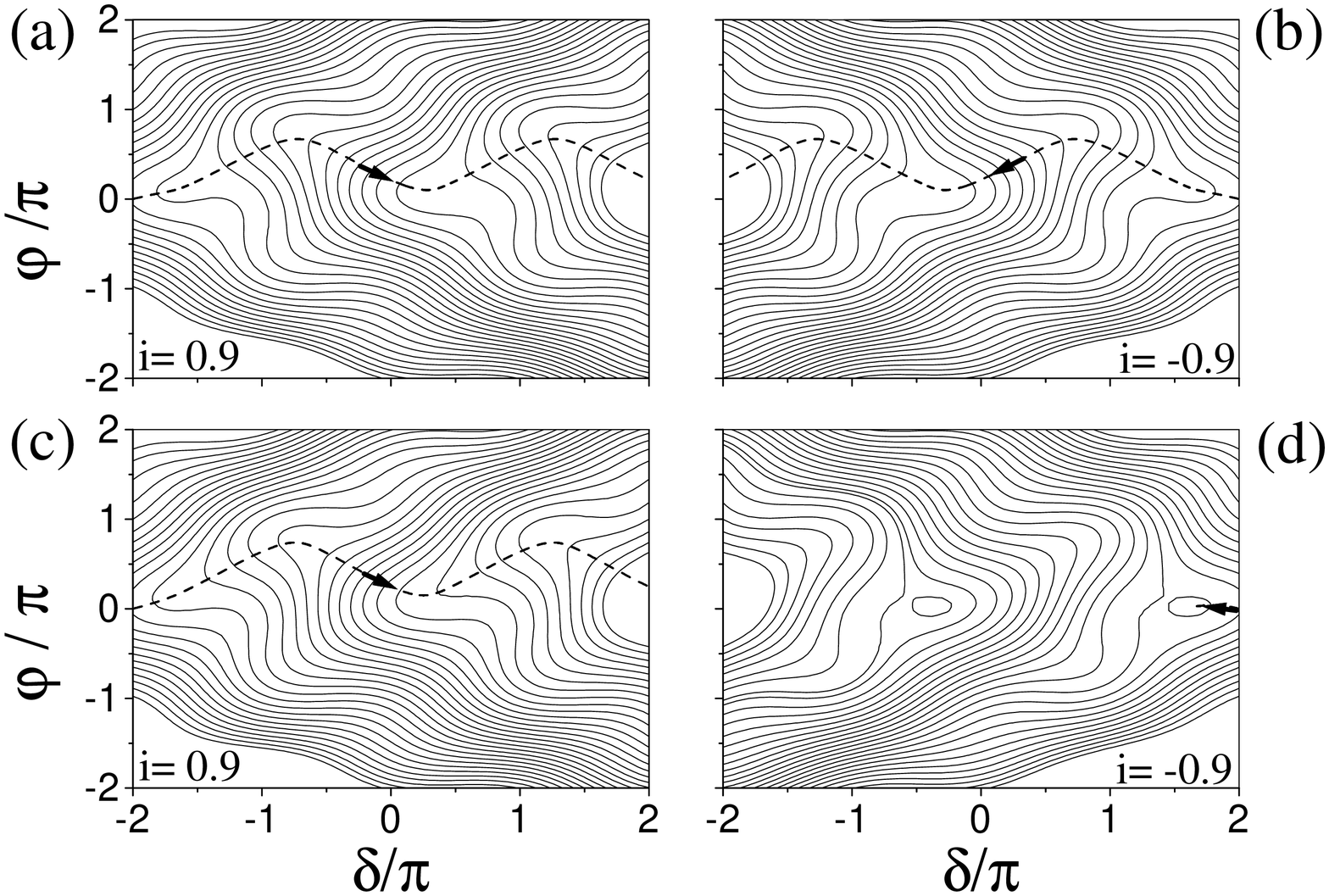}}
\refstepcounter{figure} \label{potential}
\end{figure}
Fig.~1, S. Weiss et al.
\pagebreak \thispagestyle{empty}
\begin{figure}[p]
\center{\includegraphics [width=0.5\columnwidth] {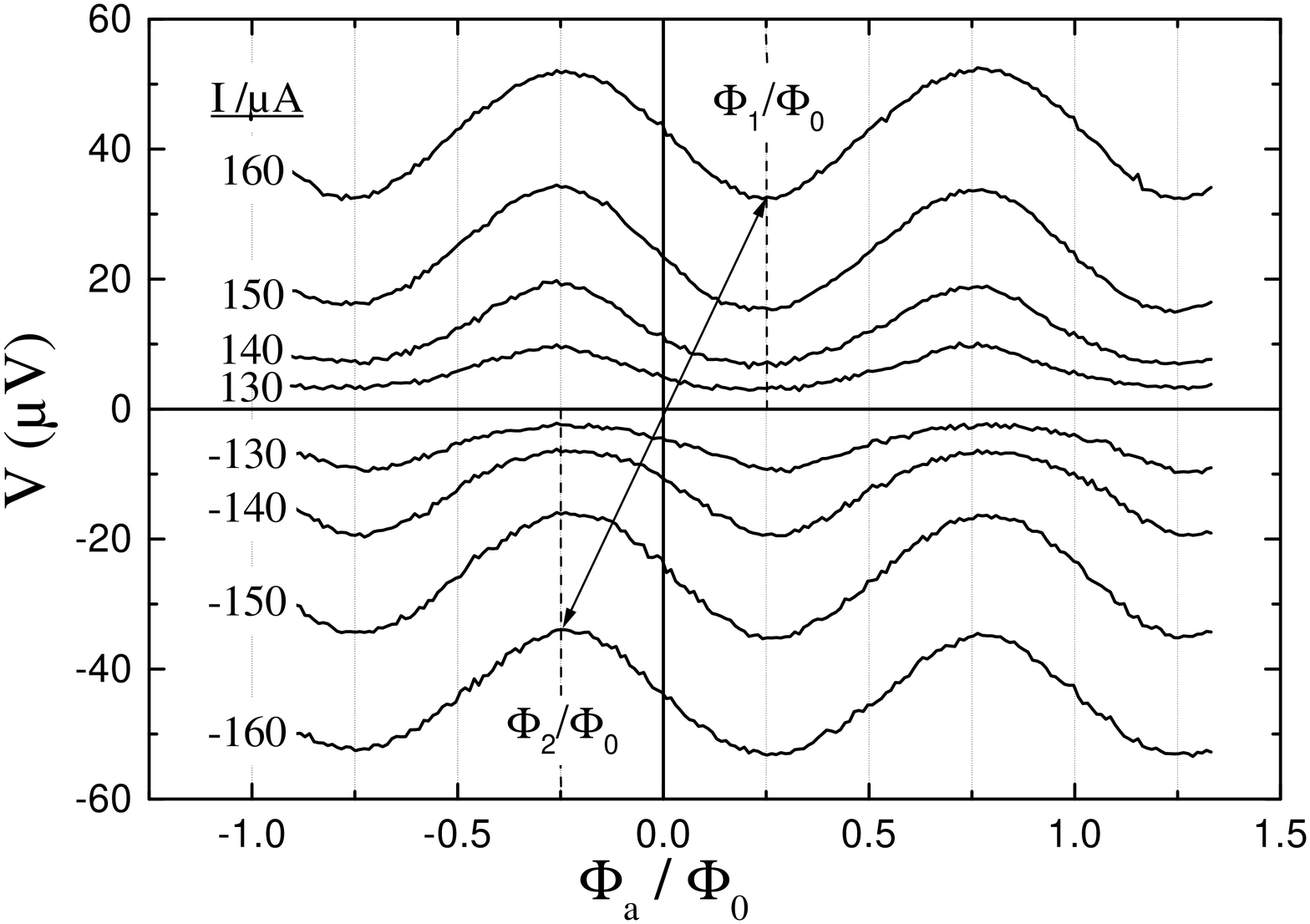}}
\refstepcounter{figure} \label{vphi}
\end{figure}
Fig.~2, S. Weiss et al.
\pagebreak \thispagestyle{empty}
\begin{figure}[p]
\center{\includegraphics [width=0.85\columnwidth] {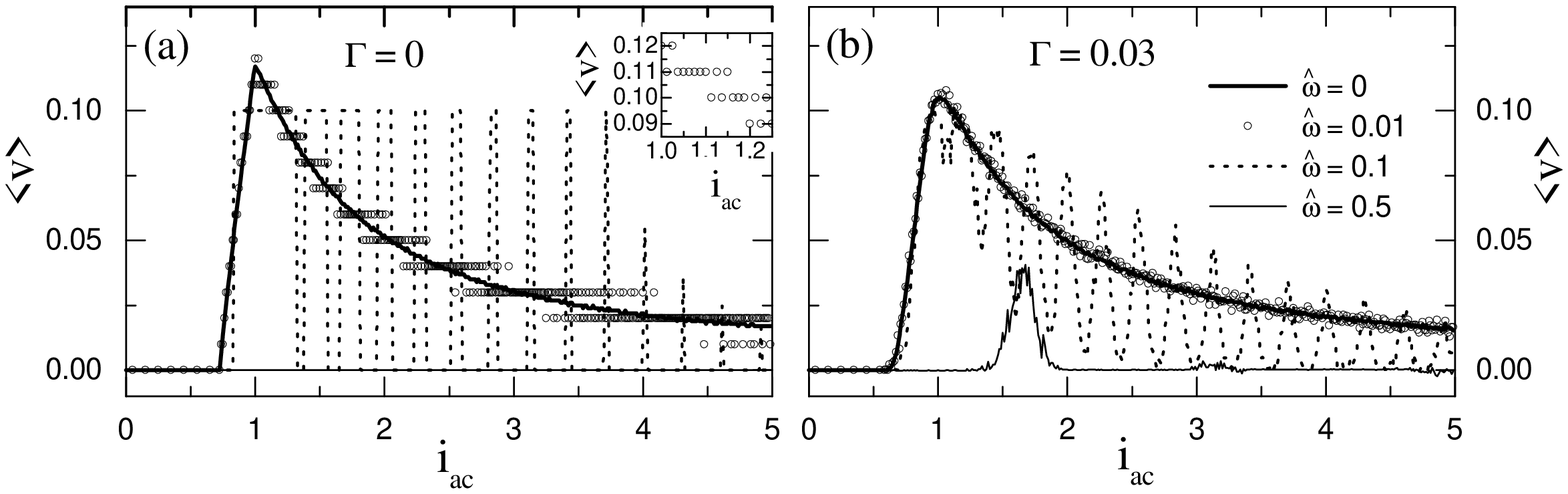}}
\refstepcounter{figure} \label{alpha_I}
\end{figure}
Fig.~3, S. Weiss et al.
\pagebreak \thispagestyle{empty}
\begin{figure}[p]
\center{\includegraphics [width=0.85\columnwidth] {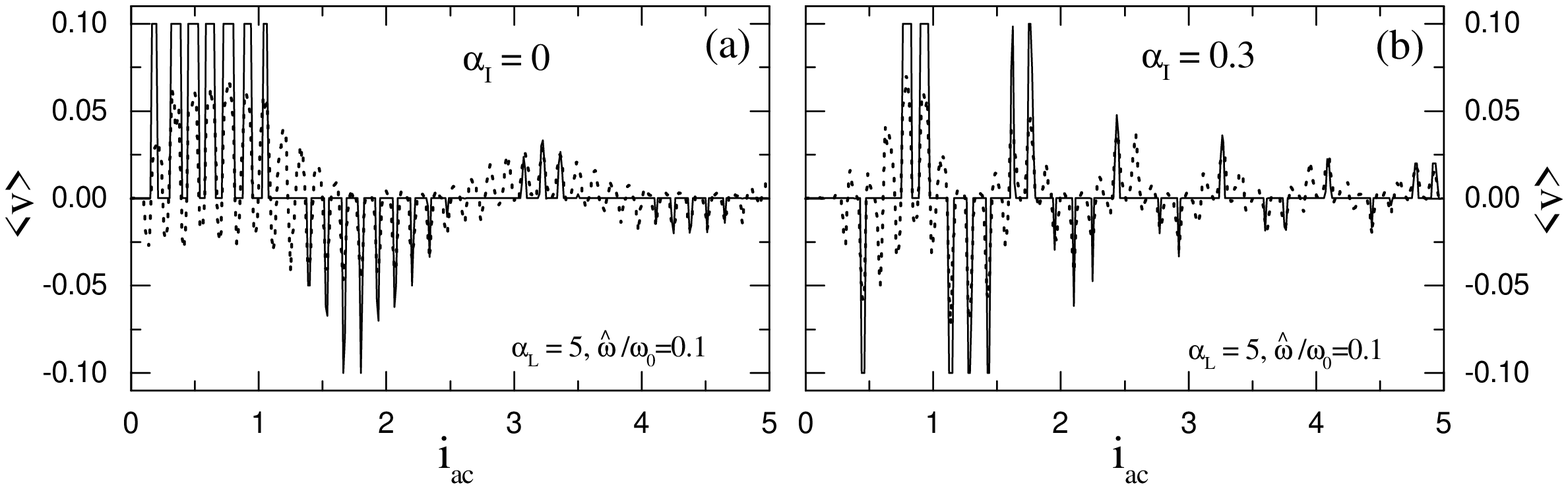}}
\refstepcounter{figure} \label{alpha_L}
\end{figure}
Fig.~4, S. Weiss et al.
\pagebreak \thispagestyle{empty}
\begin{figure}[p]
\center{\includegraphics [width=0.85\columnwidth] {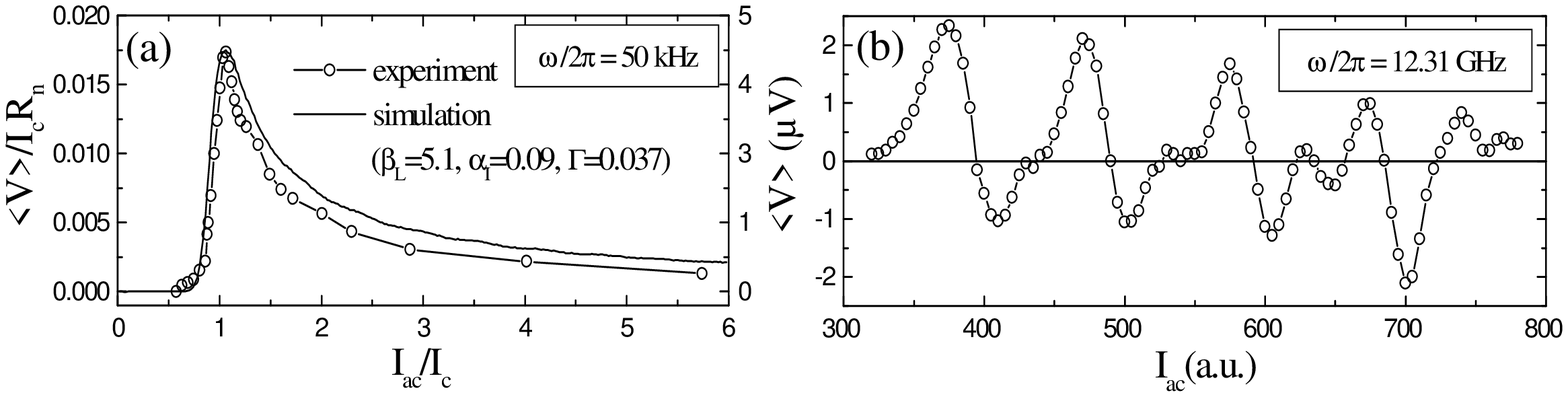}}
\refstepcounter{figure} \label{experiment}
\end{figure}
Fig.~5, S. Weiss et al.

\end{document}